\documentclass[12pt,a4]{article}
\usepackage{amsfonts,amsmath,amssymb,amscd}
\usepackage{graphicx}
\usepackage{epstopdf}

\newcommand{\beq}{\begin{equation}}
\newcommand{\eeq}{\end{equation}}

\newcommand{\kn}{| n\rangle}

\newcommand{\bn}{\langle n |}

\begin{document}
\begin{center}
{\bf{\Large 
Comparing probability distributions: application to quantum states of light}} \\
\end{center}
\begin{center}
Soumyabrata Paul$^{1, 2}$, V. Balakrishnan$^{2}$, S. Ramanan$^{1, 2}$ and S. Lakshmibala$^{2}$ \\[6pt]
$^{1}$ Department of Physics, Indian Institute of Technology Madras, Chennai 600036, India \\[6pt]
$^{2}$ Center for Quantum Information, Communication, and Computing (CQuICC), Indian Institute of Technology Madras, Chennai 600036, India \\[6pt]
\end{center}

\subsection*{Abstract}
Probability distributions play a central role in quantum mechanics, and even more so in quantum optics with its rich diversity of theoretically conceivable and experimentally accessible quantum states of light. Quantifiers that compare two different states or density matrices in terms of `distances' between the respective probability distributions include the Kullback-Leibler divergence $D_{\rm KL}$, the Bhattacharyya distance $D_{\rm B}$, and the $p$-Wasserstein distance $W_{p}$. We present a novel application of these notions to a variety of photon states, focusing particularly on the $p=1$ Wasserstein distance $W_{1}$ as it is a proper distance measure in the 
space of probability distributions. \\

\noindent
\small{{\bf Keywords}: Probability distribution, Wasserstein distance, Kullback-Leibler divergence, Bhattacharyya distance, photon number distribution}

\section{Introduction}
The light quantum or photon has played a central role in modern physics ever since its introduction by Planck and Einstein in the opening years of the $20^{\rm th}$ century. The first major step in understanding the nature of photons was the seminal work of S. N. Bose (1924), who introduced the correct photon counting statistics required to derive Planck's distribution law for the wavelength (equivalently, the frequency) of photons in blackbody radiation~\cite{Bose-1924}. We know now that blackbody radiation is an ideal quantum gas of photons (massless spin-1 bosons) obeying Bose-Einstein statistics, in thermal equilibrium at some specified temperature. \\

Major advances have been made in the last 100 years, especially since the advent of the laser as a source of coherent light. Quantum optics has come of age, and now plays an important role in quantum communication and information (see, for instance,~\cite{Fabre-2020}). From a myriad applications in technology to state-of-the-art laser interferometers in gravitational astronomy~\cite{Bailes-2021}, photonics extensively pervades modern physics. The generation of diverse quantum states of radiation, and the investigation of their statistical properties (such as the corresponding photon number distributions), are major themes of current research (see, for instance,~\cite{Browne-2017}). \\

Tomograms and pattern recognition have an important role to play in the latter case~\cite{Ibort-2009, Sharmila-2017, Laha-2018, SLVB-2022, Paul-2023}. Numerous applications of pattern recognition, image reconstruction, tomogram interpretation, and other such information-theoretic tasks require quantitative estimation of how close or how far apart two different probability distributions of a given random variable are. Among the most common quantifiers in  this regard are the Kullback-Leibler divergence~\cite{KL-1951} and the Bhattacharyya distance~\cite{Bhattacharyya-1943}, both rooted in information theory, and the Wasserstein distance or `earth-mover's distance'~\cite{Vaserstein-1969}, stemming from optimal transport theory. We shall use these quantifiers, especially the Wasserstein distance, to distinguish between different states of light arising in quantum optics~\cite{Karol-1998}.

\section{Distances between probability distributions}
Given two  normalized probability density functions (PDFs) $f(x)$ and $g(x)$ with corresponding  cumulative distribution functions (CDFs) $F(x)$ and  $G(x)$, the Kullback-Leibler divergence between them is defined as
\beq
D_{\rm KL}(f,g)
= \int_{-\infty}^{\infty} dx\, f(x)\, \mathrm{ln}\, [f(x)/g(x)].
\label{eq:KL}
\eeq
The corresponding Bhattacharyya distance is given by
\beq
D_{\rm B}(f,g)
= -\mathrm{ln} \int_{-\infty}^{\infty} dx \,
[f(x) g(x)]^{1/2}. 
\label{eq:BD}
\eeq 
$D_{\rm B}(f,g)$ is symmetric under the interchange $f\leftrightarrow g$, while $D_{\rm KL}(f,g)$ is not. Neither $D_{\rm B}$ nor $D_{\rm KL}$ is a proper distance function-- for instance, they do not satisfy the triangle inequality. The $p$-Wasserstein distance $ W_{p}(F, G)$ is defined in terms of the inverse CDFs (or quantile functions) of the probability distributions as
\beq
W_{p}(F, G) =
\Big\{\int_{0}^{1}\!dq \,
\vert F^{-1}(q) - G^{-1}(q)\vert^{p}\Big\}^{1/p},
\;\;p \geqslant 1. 
\label{eq:WD}
\eeq
It is evident that $W_{p}(F,G) = W_{p}(G,F)$. The Wasserstein distance is also a genuine distance measure in the space of probability distributions: we have 
\beq
W_{p}(F,F) = 0, \;\;
W_{p}(F,G) \geqslant 0,\;\; 
W_{p}(F,G) + W_{p}(G,H) \geqslant 
W_{p}(F,H).
\label{eq:WDproperties}
\eeq
For $p = 1$, this definition reduces (after a change of variables) to 
\beq
W_{1}(F, G) 
= \int_{-\infty}^{\infty}  dx 
\left| F(x) - G(x) \right|.
\label{eq:W1simplified}
\eeq
The modulus in the integrand is noteworthy. 
\begin{figure}
\begin{center}
\includegraphics[width=0.45\textwidth]{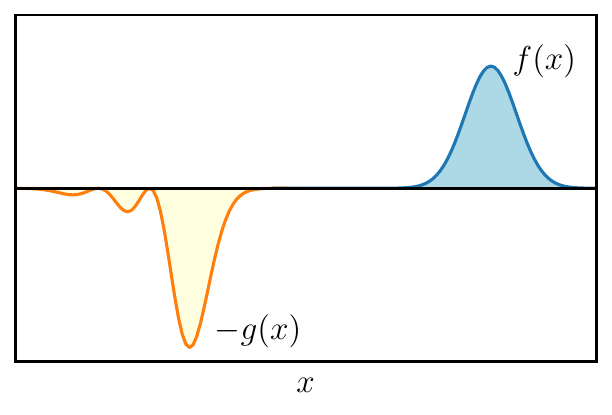}
\caption{Illustrating the earth-mover's distance in one variable.}
\label{fig_Wasserstein_schematic}
\end{center}
\end{figure}
Referring to Fig.~\ref{fig_Wasserstein_schematic}, $W_{1}(F,G)$ can be interpreted in physical terms as follows. It is the optimal cost of reshaping the `pile of earth' $f(x)$ into the pile $g(x)$. Equivalently, the optimal cost of completely flattening the pile  $f(x)$ to completely fill the pit $- g(x)$. [Cost = $\sum$ (mass $\times$ distance moved).] Hence the name ``earth-mover's distance". The counterparts of Eqs.~(\ref{eq:KL})--(\ref{eq:WD}) and Eq.~(\ref{eq:W1simplified}) for an integer-valued random variable $n$ follow straightforwardly, the integration over $x$ being replaced by a sum over $n$. While $D_{\rm KL}$ and $D_{\rm B}$ are dimensionless, $W_{1}$ has the physical dimension of the random variable concerned.

\section{A quantum mechanical example: Particle in a box}
\begin{figure}
\begin{center}
\includegraphics[width=0.45\textwidth]
{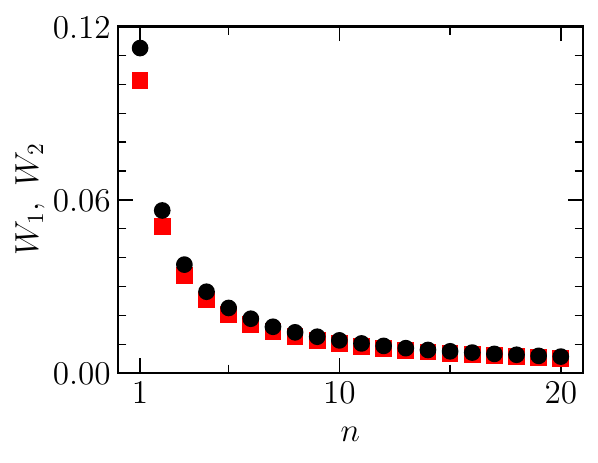}
\caption{$W_{1}$ between $G_{n}$ and the classical CDF $F_{\rm cl}(x) = x$ for the particle in a box (red squares). The $2$-Wasserstein distance $W_{2}$ is also plotted (black circles).}
\label{fig_pbox_wassdist12_gn_classical}
\end{center}
\end{figure}

The positional probability distributions of the stationary states of a particle in a 1D box $[0,1]$ provide an explicit illustration of the distances defined above. Setting the mass of the particle and $\hbar$ equal to unity, the energy levels of the particle are given by
\beq
E_n =  \tfrac{1}{2} n^{2}\pi^{2}, \;n = 1,2,
\dotsc .
\label{eq:PIABEn}
\eeq
The normalized positional PDF and CDF corresponding to $E_n$ are
\beq
g_{n} = 2 \sin^{2}(n\pi x), \quad
G_{n}(x) = x - \frac{\sin\,2n\pi x}{2n\pi}, \;
x \in [0,1].
\label{eq:PIABgn}
\eeq
Let us compare this distribution with the classical PDF in this case, which is just the uniform distribution $f_{\rm cl}(x) = 1$ with the  CDF $F_{\rm cl}(x) = x$. We find
\beq
D_{\rm B}(f_{\rm cl}, g_n) =  
\ln \, (\pi/\sqrt{8}),
\label{eq:PIABDB}
\eeq
while
\beq
D_{\rm KL}(f_{\rm cl}, g_{n}) = \ln \, 2,
\;\; D_{\rm KL}(g_{n}, f_{\rm cl})
= 1 - \ln\,2.
\label{eq:PIABKL}
\eeq
There is no $n$-dependence in these quantities. Roughly speaking, the reason is that  $f_{\rm cl}$ corresponds to the limit  $n \rightarrow \infty$, compared to which any differences in finite values of $n$ are not significant. On the other hand
\beq
W_{1}(F_{\rm cl}, G_{n}) 
= \frac{1}{2n\pi} \int_{0}^{1}
\vert \sin\,2n\pi x \vert dx = \frac{1}{n\pi^2},
\label{eq:PIABWD}
\eeq
which decreases monotonically from $1/\pi^2$ to $0$ as $n$ increases from $1$ to $\infty$, as shown in Fig.~\ref{fig_pbox_wassdist12_gn_classical}. Note that  $W_{1} (F_{\rm cl}, G_{n}) \rightarrow 0$ as $n\rightarrow \infty$, even though $g_{n} \nrightarrow f_{\rm cl}$ in this limit. While $g_{n}(x)$ continues to have a comb-like form however large  $n$ is, $f_{\rm cl}\equiv 1$. But the closeness of the two distributions is manifest in the two CDFs $G_{n}(x)$ and $F_{\rm cl}(x) = x$. \\

\begin{figure}
\begin{center}
\includegraphics[width=0.45\textwidth]
{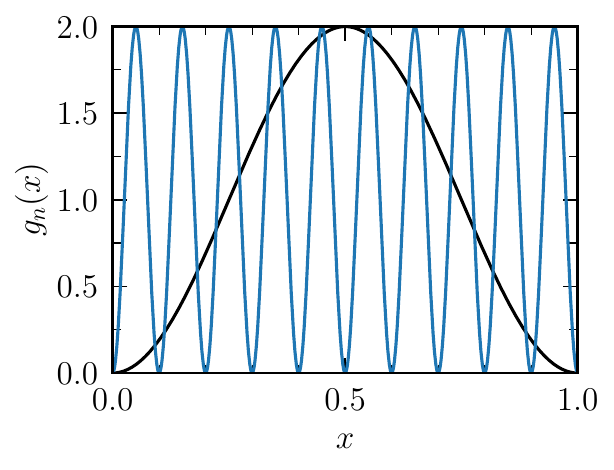}
\caption{PDFs for $n=1$ (ground state; black) and $n = 10$ (blue) for the particle in a box.}
\label{PIAB_PDF_1and10}
\end{center}
\end{figure}

\begin{figure}
\begin{center}
\includegraphics[width=0.45\textwidth]
{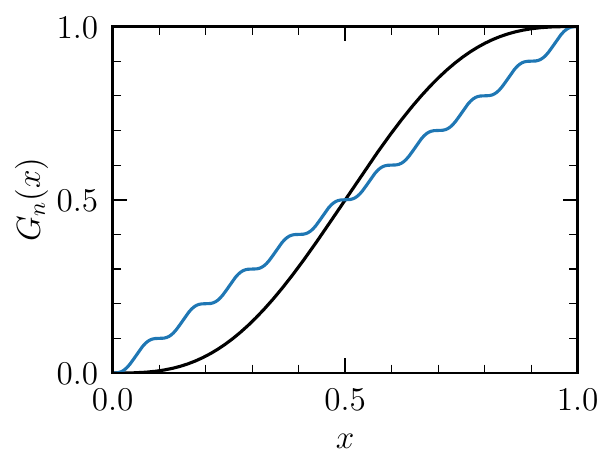}
\caption{CDFs for $n=1$ (ground state; black) and 
$n = 10$ (blue) for the particle in a box.}
\label{PIAB_CDF_1and10}
\end{center}
\end{figure}

\begin{figure}
\begin{center}
\includegraphics[width=0.45\textwidth]{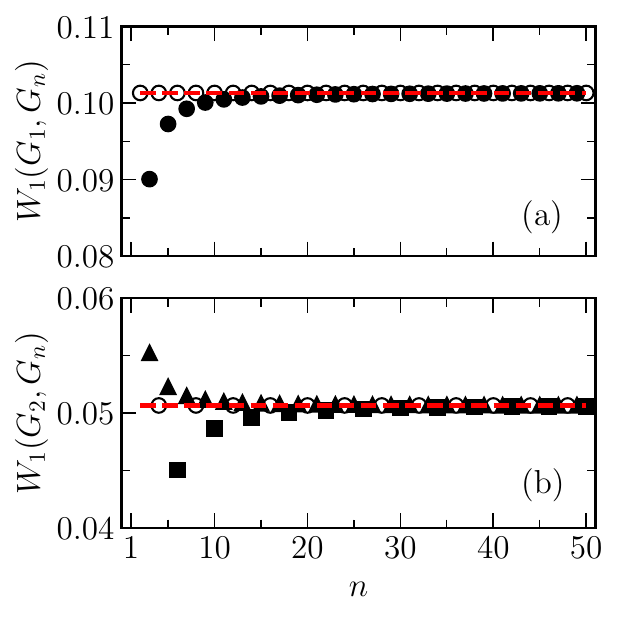}
\caption{Top to bottom: $W_{1}(G_{m}, G_{n})$ for the particle in a box versus $n$ with {\bf (a)} $m=1$ and {\bf (b)} $m=2$. {\bf (a)} Shaded circles, $n = 3,5,7\dotsc$; hollow circles, $n = 2,4,6,\dotsc$. The red dashed line denotes the value $1/\pi^{2}$. {\bf (b)} Hollow circles, $n = 4, 8, 12, \dotsc$; shaded triangles, $n = 3, 5, 7, \dotsc$; shaded squares, $n = 6, 10, 14, \dotsc$. The red dashed line denotes the value $1/({2\pi^{2}})$.}
\label{fig_pbox_wassdist1_Gn_G1_G2}
\end{center}   
\end{figure}

We may also consider $W_{1}$ between $G_{n}(x)$ and the ground state distribution $G_{1}(x)$. The PDF $g_{n} = 2 \sin^{2}(n\pi x)$ has a finer and finer comb-like form as $n$ increases, with  value $2$ at its maxima, as shown in Fig.~\ref{PIAB_PDF_1and10} for $n = 1$ and $n = 10$. 
Figure~\ref{PIAB_CDF_1and10} shows the corresponding CDFs. $W_{1}(G_{1}, G_{10})$ is the total area between the two curves in this figure.
Interestingly, there is a strong dependence of $W_{1}(G_{1}, G_{n})$ on the parity of the states concerned. For  $n = 2, 4, \dotsc$ (i.e., for all odd parity states), $W_{1}(G_{1}, G_{n}) = 1/\pi^{2}$. For $n = 3,5,\dotsc$ (even parity states), $W_{1}(G_{1}, G_{n})$ increases from $0$ and  approaches $1/\pi^{2}$ from below, as shown in Fig.~\ref{fig_pbox_wassdist1_Gn_G1_G2}(a).
Heuristically, the difference arises because the even parity PDF $g_{1}$ is unimodal with a dominant central maximum and the even PDFs $g_{3}, g_{5}, \dotsc$ also have a central peak at $x = \frac{1}{2}$. In contrast, the odd parity PDFs $g_{2}, g_{4},\dotsc$ have a central minimum, resulting in a dominant contribution to the earth-mover's distance between $G_1$ and $G_{2n}$.

\begin{figure}
\begin{center}
\includegraphics[width=0.45\textwidth]{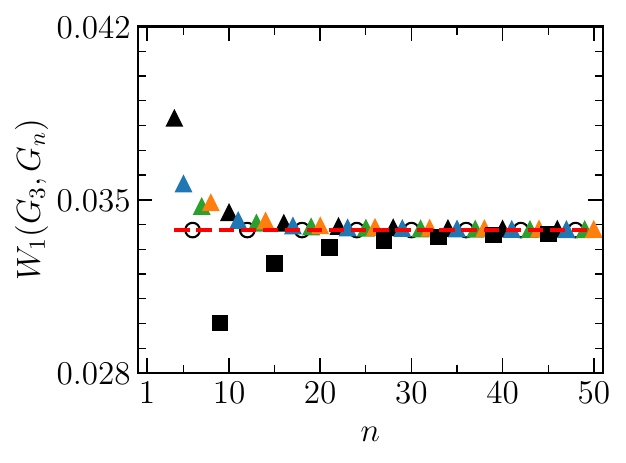}
\caption{$W_{1}(G_{3}, G_{n})$ for the particle in a box versus $n$. Hollow circles, $n=6, 12, 18, \dotsc$; black triangles, $n = 4, 10, 16, \dotsc$; blue triangles, $n=5, 11, 17, \dotsc$; green triangles, $n=7, 13, 19, \dotsc$; orange triangles, $n=8, 12, 18, \dotsc$; black squares, $n=9, 15, 21, \dotsc$. The red dashed line denotes the value $1/(3\pi^{2})$.}
\label{fig_pbox_wassdist1_Gn_G3}
\end{center}   
\end{figure}

Finer numerical effects occur when we consider $W_{1}(G_{m}, G_{n})$ for values of $m \geqslant 2$. The maxima and minima in the two PDFs intertwine in an intricate manner. Hence the heuristic parity argument does not suffice. The finer effects are seen, for instance, in a plot of $W_{1}(G_{m}, G_{n})$ versus $n$ for $m=2$ (Fig.~\ref{fig_pbox_wassdist1_Gn_G1_G2}(b)) and $m=3$ (Fig.~\ref{fig_pbox_wassdist1_Gn_G3}). Since $W_{1}(f_{\rm cl}, G_{m}) = 1/(m\pi^2)$ and $W_{1}(f_{\rm cl}, G_{n}) \longrightarrow 0$ as $n \longrightarrow \infty$, we have $W_{1}(G_{m}, G_{n}) \longrightarrow 1/(m\pi^2)$ as $n \longrightarrow \infty$.

\section{Distance between probability distributions in the photon number basis}
\subsection{Distance between a  photon number distribution and the vacuum}

We turn now to distances between states of radiation that are of interest in quantum optics. The quantifiers considered in the foregoing can be extended to discrete distributions. Let $p(n), n = 0,1,\dotsc$ be a normalized single-mode photon number distribution. $W_{1}$ between $p(n)$ and the distribution $\delta_{n,0}$ corresponding  to the vacuum state $|0\rangle$ is (by inspection) the cost of moving every $p(n)$ by the distance $n$ to that state, i.e., $\sum_{0}^{\infty} n p(n) = \overline{n}$, the mean photon number corresponding to the probability distribution $p(n)$. Similarly, it is easy to see that $W_{1}$ between the Fock states $|j\rangle$ and $|k\rangle$ is just $|j-k|$. For the displaced vacuum state or standard coherent state (CS)
\beq
|\alpha\rangle = D(\alpha)|0\rangle  
\;\;\text{where}\;\;
D(\alpha) = e^{\alpha a^{\dagger}- \alpha^{*}a}
\;(\alpha \in \mathbb{C}), 
\label{eq:displacedvac}
\eeq
we have 
\beq
W_{1}(|0\rangle, |\alpha\rangle)
=|\alpha|^{2},
\label{eq:displacedvacWD}
\eeq
the mean number of photons in the CS. This Wasserstein distance does not depend on the phase of $\alpha$. The squeezed vacuum state is given by 
\beq
|\zeta\rangle 
= S(\zeta)|0\rangle \;\;\text{where}\;\;
S(\zeta)=
e^{(\zeta^{*}a^{2}- 
\zeta a^{\dagger\,2})/2}\;
\,(\zeta \in \mathbb{C}). 
\label{squeezedvac}
\eeq
In this case we  have 
\beq
W_{1}(|0\rangle, |\zeta\rangle) = \sinh^{2}|\zeta|,
\label{eq:squeezedvacWD}
\eeq
the mean number of photons in the squeezed state. Again, this Wasserstein distance is insensitive to the phase of the complex squeezing parameter $\zeta$. \\

The corresponding Kullback-Leibler divergence $D_{\rm KL}(|0\rangle, |\alpha\rangle)$ can be calculated by noting that the PDFs in the photon number basis are given respectively by $\delta_{n,0}$ and $e^{-|\alpha|^{2}} |\alpha|^{2n}/n!$.
It follows that
\begin{equation}
D_{\rm KL}(|0\rangle, |\alpha\rangle) = |\alpha|^{2} = W_{1}(|0\rangle, |\alpha\rangle)
\label{eq:coherent_DKL}
\end{equation}
in this case.
Similarly, noting that the photon number distribution for $|\zeta\rangle$ is given by
\begin{equation}
p(n)=\frac{1}{\cosh |\zeta|} \frac{(2n)!}{2^{2n}(n!)^{2}} (\tanh |\zeta|)^{2n},
\label{eq:photon_number_dist_squeezed_vacuum}
\end{equation}
we find that 
\begin{equation}
D_{\rm KL}(|0\rangle, |\zeta\rangle) = \log \cosh |\zeta|.
\label{eq:squeezed_DKL}
\end{equation}

The examples considered above are pure states of the radiation field. As an example of a mixed state, consider any single mode of the thermal state at temperature  $T$. The  density matrix is 
\beq
\hat{\rho}_{\rm th} =
\sum_{n=0}^{\infty}p(n) |n\rangle\langle n|
\label{thermalrho}
\eeq
where the probability distribution of the number of photons of a given frequency $\nu$ and polarization is given by the geometric distribution 
\beq
p(n) = \frac{\overline{n}_{\nu}^{n}}
{(\overline{n}_{\nu}+1)^{n+1}}, \;\text{where} \;
\overline{n}_{\nu}
= \frac{1}{e^{h\nu/kT}-1}
\label{eq:thermalmean}
\eeq
is the mean number of photons corresponding to $p(n)$. Then 
\beq
W_{1}(|0\rangle\langle0|, \hat{\rho}_{\rm th})
= \overline{n}_{\nu}.
\label{eq:thermalWD}
\eeq
We further find that
\begin{equation}
D_{\rm KL}(|0\rangle\langle0|, \hat{\rho}_{\rm th}) = \overline{n}_{\nu} + 1.
\label{eq:thermalDKL}
\end{equation}
Displacing the thermal state gives another mixed state, the Glauber-Lachs state~\cite{Lachs-1965, Glauber-1966} whose density matrix is  given by 
\beq
\hat{\rho}_{\rm GL} = D(\alpha)\hat{\rho}_{\rm th}
D^{\dagger}(\alpha).
\label{eq:GLrho}
\eeq
In this case
\beq 
W_{1}(|0\rangle\langle 0|, \hat{\rho}_{\rm GL})
= |\alpha|^{2} + \overline{n}_{\nu},
\label{eq:GLWD}
\eeq
the mean number of photons in the Glauber-Lachs state. Similarly, we can calculate the Wasserstein distances between the vacuum state and many other interesting states such as the generalized CS, photon-added CS, squeezed CS, displaced squeezed states, squeezed thermal states, and so on. 

\subsection{$W_{1}$ between two photon number distributions} 

In order to generalize the results discussed in the foregoing to two non-trivial photon number distributions, it is often helpful to use the following simplification of the expression for $W_{1}$. Recall that 
\beq
W_{1}(F, G) = \int_{-\infty}^{\infty}\! \!dx\,
\vert F(x) - G(x)\vert
\label{eq:W1repeat}
\eeq
where $F$ and $G$ are the CDFs corresponding to the PDFs $f(x)$ and $g(x)$. Now suppose two conditions are satisfied: (i) $x$ only assumes non-negative values, and further (ii) $F(x) \geqslant G(x)$ for all $x\geqslant 0$. Then, since $F(\infty) = G(\infty) = 1$, we get  
\beq
W_{1}(F,G) = \int_{0}^{\infty} \!\!dx
\int_{x}^{\infty}\!\!dy  \,[g(y) - f(y)].
\label{eq:W1simplify1}
\eeq
Interchanging the order of integration, 
\beq
W_{1}(F, G) = 
\int_{0}^{\infty}\!\!dy\!
\int_{0}^{y}\!\!dx\,
\,[g(y)-f(y)]
= \overline{x}_{g}- \overline{x}_{f},
\label{eq:W1simplify2}
\eeq
the difference in the mean values of the random variable with respect to the PDFs $g(x)$ and $f(x)$. This result is also valid for a discrete random variable $n$ that takes the values $0,1,\dotsc$, such as the case of two photon number distributions $p_{1}(n)$ and $p_{2}(n)$, provided the difference $P_{1}(n) - P_{2}(n)$ of the two CDFs does not change sign for any $n$. 
Then 
\beq
W_{1}(P_{1}, P_{2}) =
\big\vert \overline{n}_{1} 
- \overline{n}_{2}\big\vert.
\label{eq:WDphotonnumber}
\eeq
Hence for two CS $|\alpha\rangle$ and $|\beta\rangle$, 
\beq
W_{1}(|\alpha\rangle, 
|\beta\rangle) = \left\vert 
|\alpha|^{2}- |\beta|^{2}\right\vert
\label{eq:WD2CS}
\eeq
in the photon number basis.

\subsection{$W_{1}$ between two blackbody frequency distributions}

As an  example of the physical significance of the Wasserstein distance, consider Planck's Law for the intensity $I(\nu)$ at any frequency $\nu$ of the radiation from a blackbody at temperature $T$, namely,
\beq
I(\nu)  \propto \frac{\nu^{3}}
{\big(e^{\beta h\nu}-1\big)},\;\, \beta = 
\frac{1}{k_{B}T}.
\label{eq:bbfreqdist}
\eeq
Normalizing $I(\nu)$ such that $\int_{0}^{\infty} I(\nu) d\nu = 1$, we may regard $I(\nu)$ as the  PDF of the random variable $\nu$, i.e., the PDF for any photon in the photon gas at energy $h\nu$. Let $I_1(\nu)$ and $I_2(\nu)$ be the normalized PDFs of the frequency for two blackbodies at respective temperatures $T_1$ and $T_2$, and let $\overline{\nu}_1$ and $\overline{\nu}_2$ denote the respective mean values of the frequency in the two cases. Then $W_{1}(I_1, I_2)=\vert \overline{\nu}_1 - \overline{\nu}_2\vert$. But the only frequency scale in the PDF $I(\nu)$ is $k_{B}T/h$. Hence the mean frequency $\overline{\nu}$ must necessarily be a constant multiple of $k_{B}T/h$, so that
\beq
W_{1}(I_1, I_2) \propto \vert T_1 - T_2\vert.
\label{eq:bbT1T2}
\eeq
Similarly, if $\mathcal{J}(\lambda)$ denotes the normalized PDF of the wavelength of the photons in the radiation, an analogous argument ($ch/k_{B}T$ is the only length scale in $\mathcal{J}(\lambda)$) yields
\beq
W_{1}(\mathcal{J}_1, \mathcal{J}_2)
\propto \vert T_{1}^{-1} - T_{2}^{-1}\vert.
\label{eq:bblambda}
\eeq

\section{Distance between photon number states in the 
$x$-representation}

\begin{figure}
\begin{center}
\includegraphics[width=0.45\textwidth]
{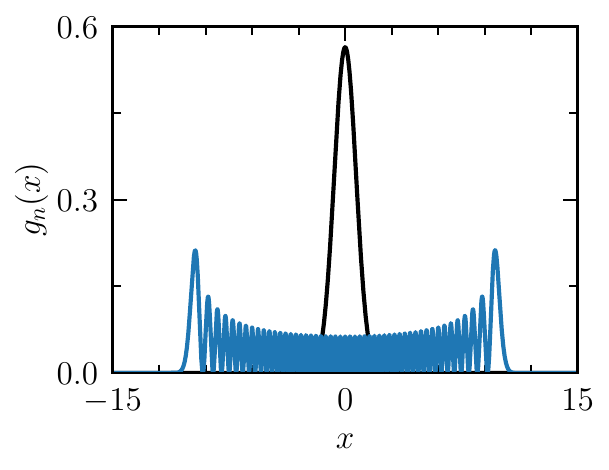}
\caption{PDFs in $x$ for the $0$-photon state (black) and the $50$-photon state (blue).}
\label{fig_HO_psiSqr_n_0_50}
\end{center}
\end{figure}

The PDFs under discussion, and hence the quantifiers of the distance between two states, are obviously both basis-dependent and representation-dependent. Consider, for instance,  the Fock states $\kn$ and $|0\rangle$ in the $x$-quadrature $(\hat{a}+ \hat{a}^{\dagger})/\sqrt{2}$, where $\hat{a}, \hat{a}^{\dagger}$ are the photon annihilation and creation operators 
of the single-mode radiation. The density matrix in the $x$-representation is 
\beq
\langle x\kn\bn x \rangle =
g_n(x) = \frac{1}{2^{n} \sqrt{\pi} n!}\, e^{-x^{2}} 
H_{n}^2(x),
\label{eq:gnxrepresentation}
\eeq 
where $H_{n}$ is the Hermite polynomial of order $n$. Figure~\ref{fig_HO_psiSqr_n_0_50} compares the $0$-photon (or vacuum state) PDF $g_{0}(x)$  and the $50$-photon PDF $g_{50}(x)$. The former is unimodal, with a peak at $x = 0$. The latter is multimodal, with the most prominent peaks at its edges, owing to the leading large-$|x|$ behavior ($\sim x^{2n}$) of $H_{n}^{2}(x)$. As $n$ increases, this trend gets more pronounced, and the envelope of $g_{n}(x)$ tends to the shape of the positional PDF of a classical oscillator. \\

\begin{figure}
\begin{center}
\includegraphics[width=0.45\textwidth]
{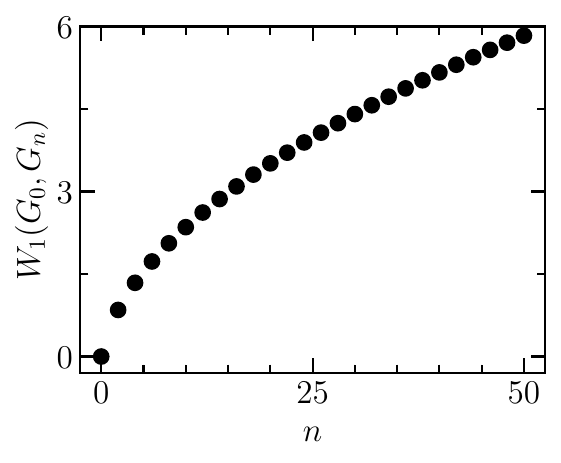}
\caption{$W_{1}(G_0, G_{n})$ versus $n$ in the $x$-quadrature for the states $|0\rangle$ and $\kn$.}
\label{fig_sho_wassdist1}
\end{center}
\end{figure}

It is instructive to compute $W_{1}$ between the $0$-photon state and the $n$-photon state from their respective CDFs in the $x$-quadrature, namely, the expression 
\beq
W_{1}(G_0, G_{n})
 = \int_{-\infty}^{\infty}\!
dx\,\vert G_{0}(x) - G_{n}(x)\vert.
\label{eq:G0Gnxquad}
\eeq
Figure~\ref{fig_sho_wassdist1} depicts $W_{1}(G_0, G_{n})$ as a function of the photon number $n$.
This $W_{1}$ has a leading large-$n$ behavior $\sim n^{1/2}$, which is expected on physical grounds. As already noted, in the state $|0\rangle$, most of the probability mass in $g_{0}(x)$ is concentrated around $x = 0$. On the other hand, in the state $\kn$, most of the probability mass accumulates for large $n$ around $\pm E_{n}^{1/2} \sim \pm n^{1/2}$, (the classical turning points), for both even and odd $n$. This feature becomes more pronounced with increasing $n$. The earth-mover's distance from $g_{0}(x)$ $g_{n}(x)$ is therefore $ \sim  n^{1/2}$ asymptotically. \\
 
\begin{figure}
\begin{center}
\includegraphics[width=0.45\textwidth]
{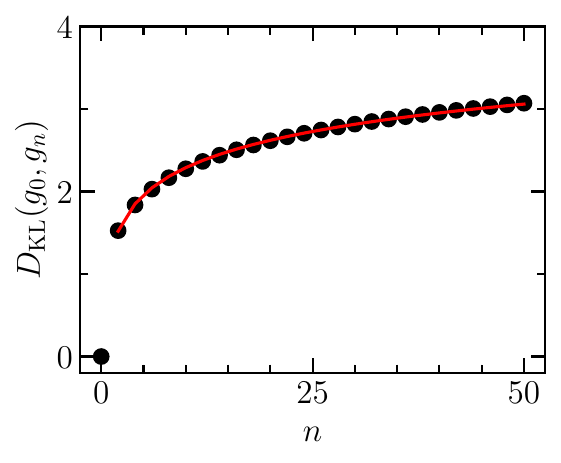}
\caption{$D_{\rm KL}(g_0, g_n)$ versus $n$. The red line is a fit to a linear function of $\ln\,n$.}
\label{fig_sho_KL} 
\end{center}
\end{figure}

\begin{figure}
\begin{center}
\includegraphics[width=0.45\textwidth]{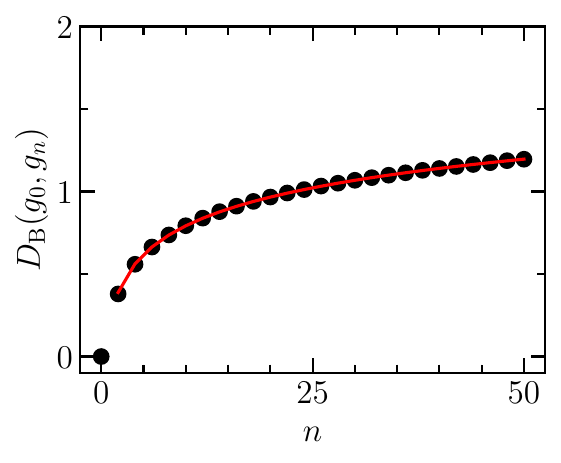}
\caption{$D_{\rm B}(g_0, g_n)$ versus $n$. The red line is a fit to a linear function of $\ln\,n$.}
\label{fig_sho_BD}
\end{center}
\end{figure}

This behavior of $W_{1}(G_0, G_{n})$ is to be compared with that of the corresponding $D_{\rm KL}$ and $D_{\rm B}$. In this case, the Kullback-Leibler divergence reduces to
\beq
D_{\rm KL}(g_0, g_n)
= \ln \,(2^n n!) - \frac{2}{\sqrt{\pi}} \int_0^\infty  dx\, e^{-x^2} \, \ln\, 
H_n^2(x).
\label{eq:KLosc}
\eeq
The variation of this expression as a function of $n$ is shown in Figure~\ref{fig_sho_KL}.  
Likewise, the corresponding Bhattacharyya distance is given by
\beq
D_{\rm B}(g_0, g_n) 
= -\ln \, \Big\{\frac{2}{\sqrt{2^n \, n! \, \pi}}\, \int_0^\infty  dx \, e^{-x^2} \, |H_n(x)| \Big\}.
\label{eq:Bhattosc}
\eeq
Figure~\ref{fig_sho_BD} shows the variation of $D_{\rm B}(g_0, g_n)$ with $n$. 
We can establish that the asymptotic behavior of both $D_{\rm KL}$ and $D_{\rm B}$ is of the form $a\,\ln\,n + b$, in contrast to the $n^{1/2}$ behavior of $W_{1}$. Hence $W_{1}$ is a much more sensitive quantifier than $D_{\rm KL}$ or $D_{\rm B}$ in the present instance. \\ 

\section{Concluding remarks}
The Wasserstein distance, the Kullback-Leibler divergence and the Bhattacharyya distance are quantitative estimates of the degree of closeness of different probability distributions pertaining to a given system. These quantifiers provide useful information in the comparison of the respective photon number distributions in appropriate pairs of quantum states of radiation. This aspect blends well with the analysis of states in quantum optics, especially time-evolving states, using a tomographic approach that relies only on probability distributions and avoids tedious state reconstruction. This feature makes machine learning protocols based on the Wasserstein distance and the Kullback-Leibler divergence for distinguishing between different states of light very advantageous in practice. The details of such applications will be presented elsewhere. \\

\noindent
{\bf Acknowledgments}: \\
We acknowledge partial support through funds from Mphasis to the Centre for Quantum Information, Communication and Computing (CQuICC), IIT Madras. VB and SL thank the Department of Physics, Indian Institute of Technology Madras for infrastructural support. \\

\noindent
{\bf Conflict of interest statement}: \\
On behalf of all authors, the corresponding author states that there is no conflict of interest.


\noindent
\begin{thebibliography}{99}
\bibitem{Bose-1924} S.~N.~Bose, Z. Physik \textbf{26}, 178 (1924).
\bibitem{Fabre-2020} C.~Fabre and N.~Terps, Rev. Mod. Phys. \textbf{92}, 035005 (2020).
\bibitem{Bailes-2021} M.~Bailes \textit{et al.}, Nat. Rev. Phys. \textbf{3}, 344 (2021).
\bibitem{Browne-2017} D.~Browne, S.~Bose, F.~Mintert and M.~S.~Kim, Prog. Quantum Electron. \textbf{54}, 2 (2017).
\bibitem{Ibort-2009} A.~Ibort, V.~I.~Man'ko, G.~Marmo, A.~Simoni and F.~Ventriglia, Phys. Scr. \textbf{79}, 065013 (2009).
\bibitem{Sharmila-2017} B.~Sharmila, K.~Saumitran, S.~Lakshmibala and V.~Balakrishnan, J. Phys. B: At. Mol. Opt. Phys. \textbf{50}, 045501 (2017).
\bibitem{Laha-2018} P.~Laha, S.~Lakshmibala and V.~Balakrishnan, J. Mod. Opt. \textbf{65}, 1466 (2018).
\bibitem{SLVB-2022} S.~Lakshmibala and V.~Balakrishnan, \textit{Nonclassical Effects and Dynamics of Quantum Observables}, Springer Briefs in Physics, Springer (2022).
\bibitem{Paul-2023} S.~Paul, S.~Lakshmibala, V.~Balakrishnan and S.~Ramanan, Front. Quantum Sci. Technol. \textbf{2}, 1131798 (2023).
\bibitem{KL-1951}  R.~Kullback and R.~A.~Leibler, Ann. Math. Statist. \textbf{22}, 79 (1951).
\bibitem{Bhattacharyya-1943} A.~Bhattacharyya, Bull. Cal. Math. Soc. \textbf{35}, 99 (1943).
\bibitem{Vaserstein-1969} L.~N.~Vaserstein, Probl. Peredachi Inf. \textbf{5}, 64 (1969).
\bibitem{Karol-1998} The Wasserstein distance between the Husimi quasiprobability distributions corresponding to some states of interest in quantum optics has been considered in K. Zyczkowski and W. Slomczynski, J. Phys. A: Math. Gen. \textbf{31}, 9095 (1998).
\bibitem{Lachs-1965} G.~Lachs, Phys. Rev. \textbf{138}, B1012 (1965).
\bibitem{Glauber-1966} R.~J.~Glauber, in \textit{Progress of Quantum Electronics}, 
Eds.  P.~L.~Kelley, B.~Lax and P.~E.~Tannenwald, McGraw-Hill (1966).
\end{thebibliography}
\end{document}